\documentclass[final,5p,times,twocolumn]{elsarticle}

\usepackage{amssymb}
\usepackage{lineno}
\usepackage[T1]{fontenc}

\journal{Nuclear Instruments \& Methods in Physics Research, Section A}

\begin{document}


\begin{frontmatter}

\title{The PANDA Barrel DIRC}

\author[1]{R.~Dzhygadlo\corref{cor1}}
\ead{r.dzhygadlo@gsi.de}

\cortext[cor1]{Corresponding author}

\author[1]{A.~Belias}
\author[1]{A.~Gerhardt}
\author[1]{D.~Lehmann}
\author[1,2]{K.~Peters}
\author[1]{G.~Schepers}
\author[1]{C.~Schwarz}
\author[1]{J.~Schwiening}
\author[1]{M.~Traxler}
\author[1,2]{Y.~Wolf}
\author[3]{L.~Schmitt}
\author[4]{M.~B\"{o}hm}
\author[4]{K.~Gumbert}
\author[4]{S.~Krauss}
\author[4]{A.~Lehmann}
\author[4]{D.~Miehling}
\author[5]{M.~D\"{u}ren}
\author[5]{A.~Hayrapetyan}
\author[5]{I.~K\"{o}seoglu}
\author[5]{M.~Schmidt}
\author[5]{T.~Wasem}
\author[6]{C.~Sfienti}
\author[7]{A.~Ali}

\affiliation[1]{organization={GSI Helmholtzzentrum f\"{u}r Schwerionenforschung GmbH},
  city={Darmstadt}, country={Germany}}
\affiliation[2]{organization={Goethe University},
  city={Frankfurt}, country={Germany}}
\affiliation[3]{organization={FAIR, Facility for Antiproton and Ion Research in Europe},
  city={Darmstadt}, country={Germany}}
\affiliation[4]{organization={Friedrich Alexander-University of Erlangen-Nuremberg},
  city={Erlangen}, country={Germany}}
\affiliation[5]{organization={II. Physikalisches Institut, Justus Liebig-University of Giessen},
  city={Giessen}, country={Germany}}
\affiliation[6]{organization={Institut f\"{u}r Kernphysik, Johannes Gutenberg-University of Mainz},
  city={Mainz}, country={Germany}}
\affiliation[7]{organization={Helmholtz-Institut Mainz},
  city={Mainz}, country={Germany}}

\begin{abstract}

  The PANDA experiment at the international accelerator Facility for Antiproton and Ion Research in Europe (FAIR), Darmstadt, Germany, will address fundamental questions of hadron physics using $\bar{p}p$ annihilations. Excellent Particle Identification (PID) over a large range of solid angles and particle momenta will be essential to meet the objectives of the rich physics program. Charged PID in the target region will be provided by a Barrel DIRC (Detection of Internally Reflected Cherenkov light) counter.
  The Barrel DIRC, covering the polar angle range of 22-140 degrees, will provide a $\pi/K$ separation power of at least 3 standard deviations for charged particle momenta up to 3.5 GeV/c. The design of the Barrel DIRC features narrow radiator bars made from synthetic fused silica, an innovative multi-layer spherical lens focusing system, a prism-shaped synthetic fused silica expansion volume, and an array of lifetime-enhanced Microchannel Plate PMTs (MCP-PMTs) to detect the hit location and arrival time of the Cherenkov photons. Detailed Monte-Carlo simulations were performed, and reconstruction methods were developed to study the performance of the system.
  All critical aspects of the design and the performance were validated with system prototypes in a mixed hadron beam at the CERN PS. In 2020 the PANDA Barrel DIRC project advanced from the design stage to component fabrication.
  The series production of the fused silica bars was successfully completed in 2021 and delivery of the MCP-PMTs started in May 2022.

\end{abstract}


\end{frontmatter}

\section{Introduction}

\begin{figure}[ht]
  \centering
  \includegraphics[width=0.95\columnwidth]{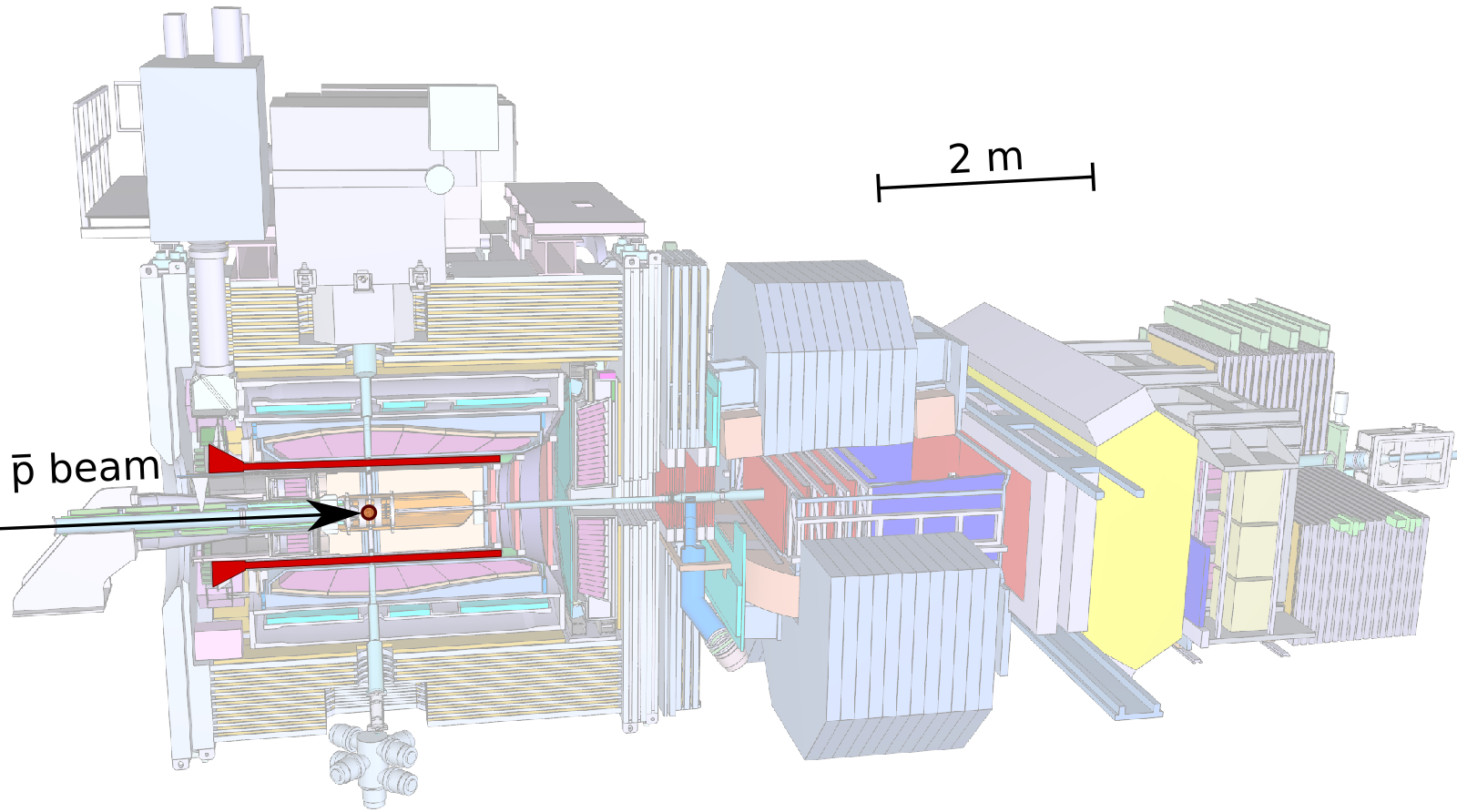}
  \caption{Schematic cross-section of the PANDA detector.
    The Barrel DIRC is located in the target spectrometer region and is highlighted with red color.}
  \label{panda_det}
\end{figure}

The PANDA experiment \cite{panda-physics} is designed to address
fundamental questions of hadron physics using high-intensity
cooled antiproton beams with momenta between
1.5 and 15~GeV/$c$ and a design luminosity of up to $2 \times 10^{32} $cm$^{-2} $s$^{-1}$.
A sophisticated detector system (see Fig.~\ref{panda_det}) with an acceptance close to 4$\pi$,
precise tracking, calorimetry, and particle identification (PID) was designed to accomplish that goal \cite{panda1}.
The hadronic PID in the target region will be provided by two Cherenkov detectors based on
the DIRC (Detection of Internally Reflected Cherenkov light) principle.
The goal for Barrel DIRC is to separate kaons and pions with at least 3 standard deviations (s.d.) for momenta up
to 3.5 GeV/c and polar angles between 22$^\circ$ and 140$^\circ$.
The PID in the forward region from 5$^\circ$ to 22$^\circ$ will be provided by the Endcap Disk DIRC \cite{endcap}.

\section{Detector design and performance}

\begin{figure}[ht]
  \centering
  \includegraphics[width=0.95\columnwidth]{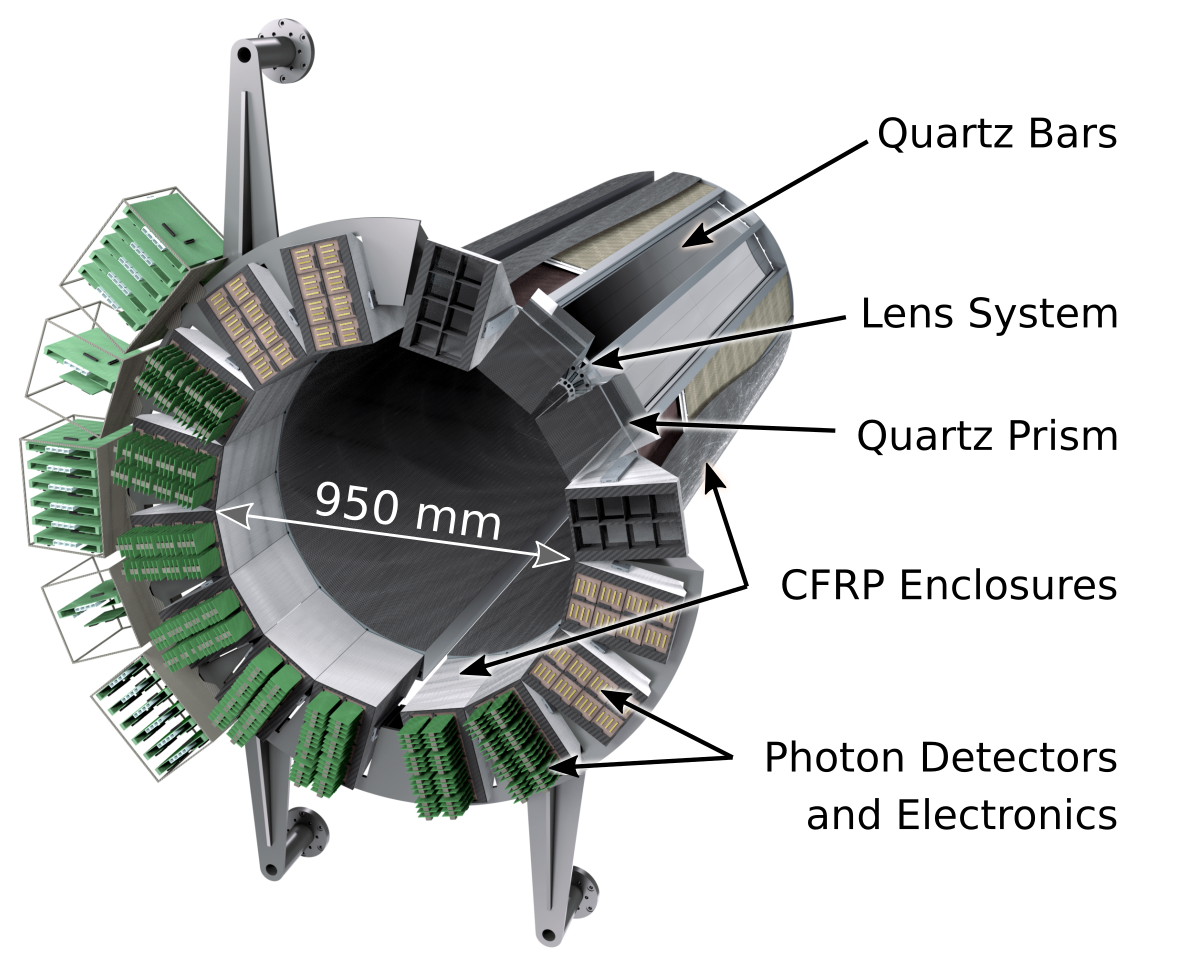}

  \vspace{-0.12cm}
  
  \includegraphics[width=0.9\columnwidth]{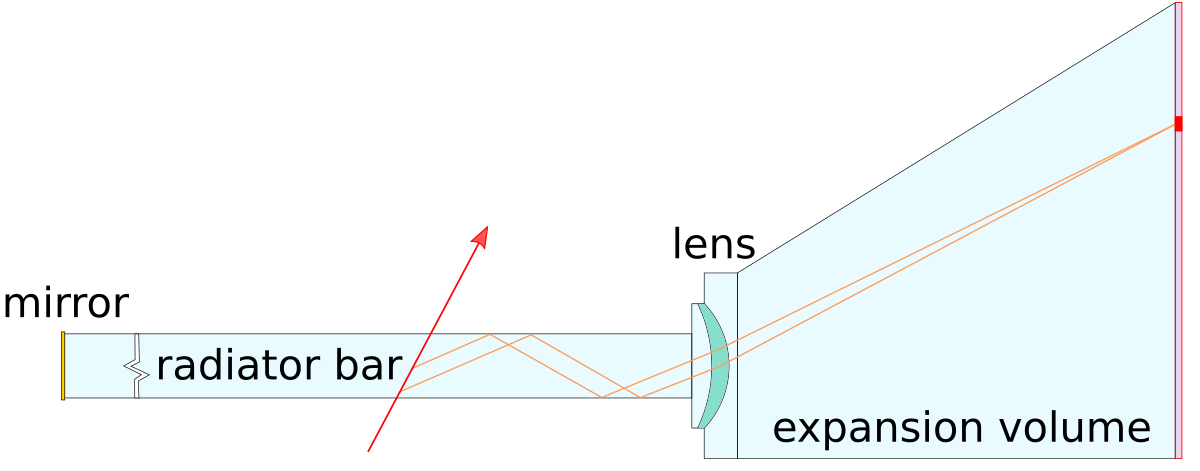}
  \caption{Rendered CAD drawing of the PANDA Barrel DIRC (top)
    and schematic drawing of the photon path inside the DIRC (bottom, not to scale). }
  \label{design_lh}
\end{figure}

The design of the PANDA Barrel DIRC \cite{tdr} (see Fig.~\ref{design_lh})
is inspired by the BaBar DIRC counter \cite{adam2005}.
It is constructed in the form of a barrel using 16 optically isolated sectors, each comprising
a bar box and a compact, prism-shaped expansion volume (EV).
Each bar box is about 167~mm wide and contains three synthetic fused silica bars of 17 $\times$ 53 $\times$ 2400~mm$^3$ size, placed side-by-side, separated by 0.1~mm-wide air gaps to ensure total internal reflection.
A flat mirror at the forward end of each bar is used to reflect
the Cherenkov photons to the read-out end, where a 3-layer 
spherical lens images them on an array of 8 Microchannel Plate
Photomultiplier Tubes (MCP-PMTs).
Each MCP-PMT has 64 pixels of 6.5 $\times$ 6.5~mm$^2$ size
and, in combination with the FPGA-based readout electronics, 
will be able to detect single photons with a precision of about $100$~ps.

\begin{figure}[ht]
  \centering
  \includegraphics[width=0.9\columnwidth]{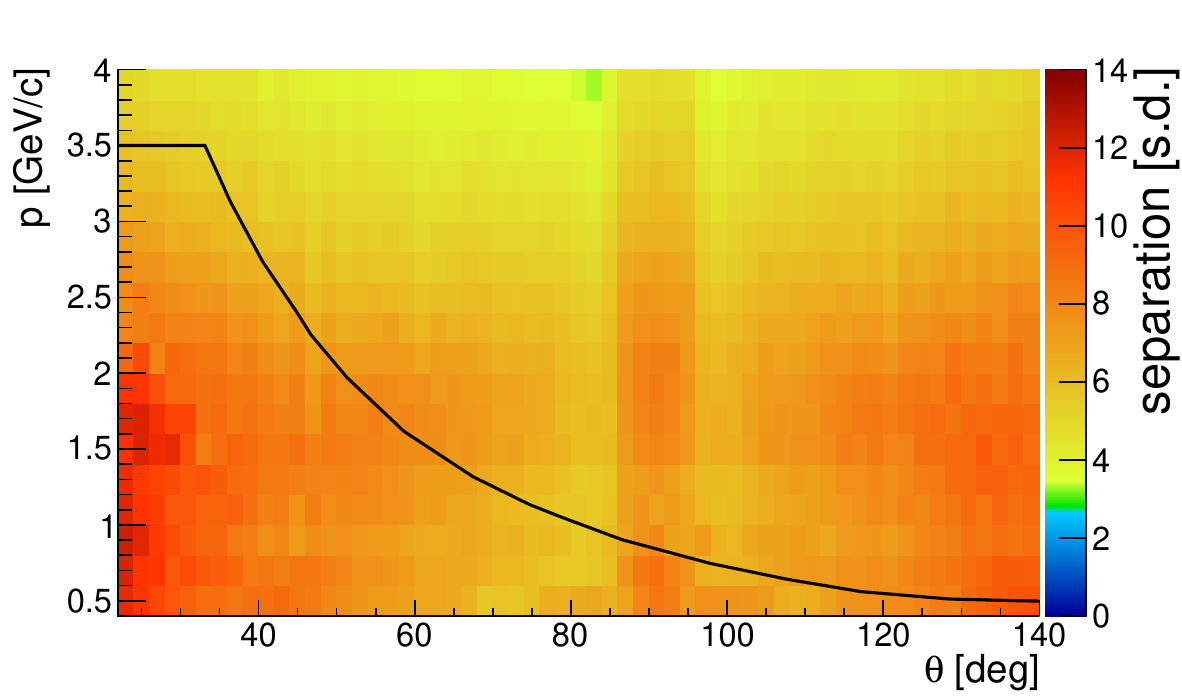}
  \caption{$\pi/K$ separation power as a function of particle momentum and polar angle.
    The black line outlines expected phase space distribution of kaons at $p_{\bar{p}}=7$ GeV/c
    and the PID requirement of the Barrel DIRC.}
  \label{sep_map}
\end{figure}

A detailed Monte Carlo simulation based on Geant4 \cite{geant4} was developed
within the PandaRoot framework \cite{pandaroot} to determine the expected PID performance of the system. 
The parameters of the simulation were tuned to the
experimentally measured values for all relevant quantities, such as
quantum and collection efficiency, gain uniformity, timing resolution of the MCP-PMTs,
and characteristics of the optical materials.
Two reconstruction methods where used to benchmark the performance of the design \cite{RICH14_sim}.
The \textit{geometrical reconstruction}, initially used by the BaBar DIRC,
performs PID by reconstructing the value of the Cherenkov angle
and using it in a track-by-track maximum likelihood fit.
This method relies mostly on the position of the detected photons in the reconstruction,
while the \textit{time imaging} utilizes both, position and time information,
and directly performs the maximum likelihood fit.
The time imaging method was initially developed for the Belle II time-of-propagation (TOP)
counter \cite{staric_2} and further adapted for the PANDA Barrel DIRC \cite{tireco}.
The PID performance of the time imaging reconstruction of the Geant4 simulations is represented in
Fig.~\ref{sep_map} where the $\pi/K$ separation power is shown as a function of the particle
momentum and polar angle.
A separation power of 4 s.d. or more is expected for the entire $\pi/K$ phase space,
exceeding the PANDA PID requirements.

\section{Prototype tests}

Most core design aspects were tested in particle beams at the CERN PS \cite{schwarz2020}.
The latest campaign at CERN in August 2018 led to the final design validation
for the Barrel DIRC with a mixed hadronic beam.
The prototype of the Barrel DIRC was placed into the beamline together with supplementary detectors,
which included a set of scintillators for DAQ triggering and a scintillating fiber hodoscope
for confining the beam within a few millimeter radius. In addition, a fast time-of-flight system
was used to provide external PID information.
The prototype support frame could be translated manually and rotated remotely
relative to the beam, making it possible to scan the equivalent of the PANDA Barrel DIRC phase space.

\begin{figure}[ht]
  \centering
  \includegraphics[width=0.9\columnwidth]{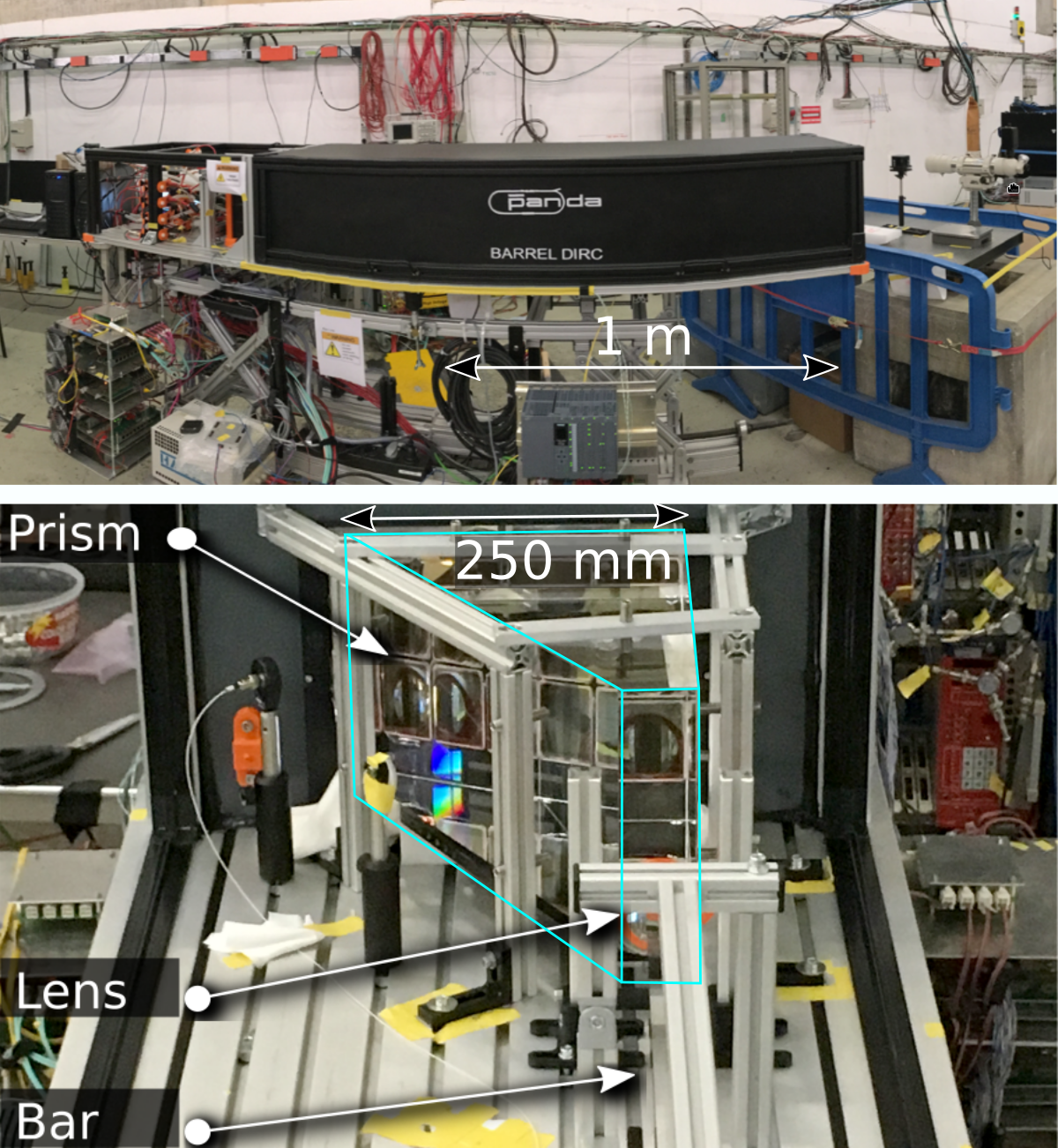}
  \caption{Photographs of the Barrel DIRC prototype in the CERN PS T9 beamline (top)
    and its main components (bottom).}
  \label{prototype_beam}
\end{figure}

Fig.~\ref{prototype_beam} shows photographs of the Barrel DIRC prototype in the T9 beamline
at the CERN PS (top) and its main components (bottom).
The prototype included the key elements of one
PANDA Barrel DIRC sector. A narrow fused silica bar
(17.1 $\times$ 34.9 $\times$ 1200.0~mm$^3$) was used as radiator.
It was coupled on one end to a flat mirror and on the other end to a 3-layer spherical
focusing lens with a fused silica prism as expansion volume.
An array of 2 $\times$ 4 PHOTONIS Planacon XP85012/A1 MCP-PMTs \cite{pmt}, attached to the back side of
the EV, was used to detect the Cherenkov photons.
In total 512 electronics channels were read out using Trigger and Readout Boards
(TRB) \cite{trb} in combination with
FPGA-based amplification and discrimination cards (PADIWA) \cite{cardinali}.
The timing calibration was performed by using a picosecond laser pulser \cite{pilas}
which produces pulses with FWHM of 27~ps at 372~nm wavelength. The laser was coupled through
optical fibers to two diffusers for illumination the entire photo-detection plane.
The average timing precision per channel achieved during the beam test was measured to be $\sigma \approx 200$~ps.

The direct measurements of the $\pi/K$ separation power at the targeted momentum of 3.5~GeV/$c$
was not possible since the T9 beam was composed mostly of $e$, $\mu$, $\pi$ and $p$.
Instead, the PID performance was evaluated for $\pi$ and $p$ at 7~GeV/$c$,
where the Cherenkov angle difference is equivalent to $\pi$ and $K$ at 3.5~GeV/$c$.

\begin{figure}[ht]
  \centering
  \includegraphics[width=0.9\columnwidth]{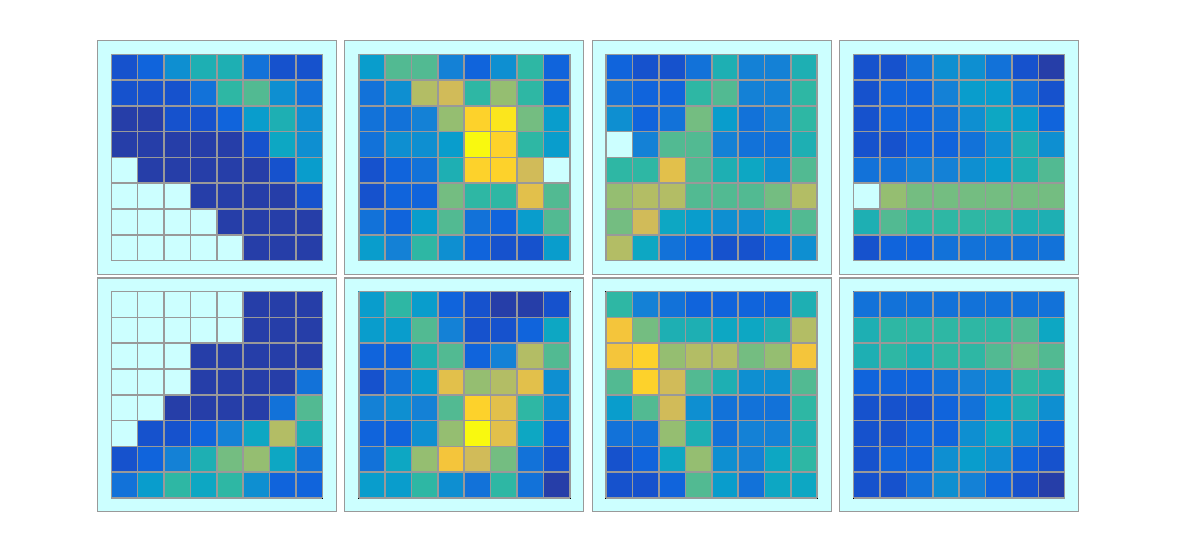}

  \vspace{-0.2cm}

  \includegraphics[width=0.9\columnwidth]{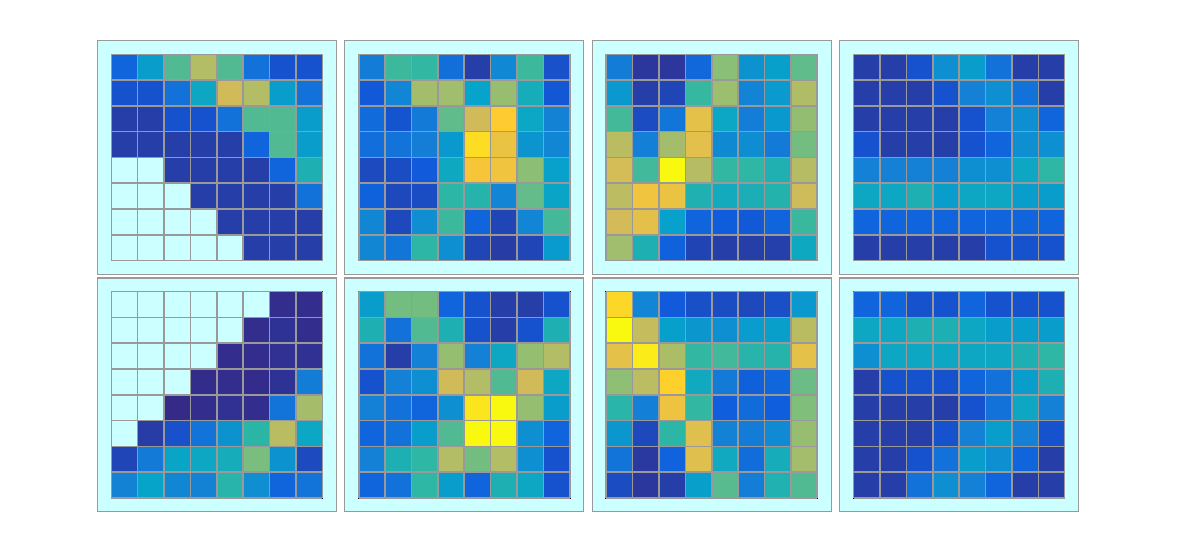}
  \caption{Accumulated hit pattern of 2000 pions at 7~GeV/$c$ momentum and $20^\circ$ polar angle from beam data (top) and simulation (bottom).}
  \label{hp}
\end{figure}

\begin{figure}[ht]
  \centering
  \includegraphics[width=0.9\columnwidth]{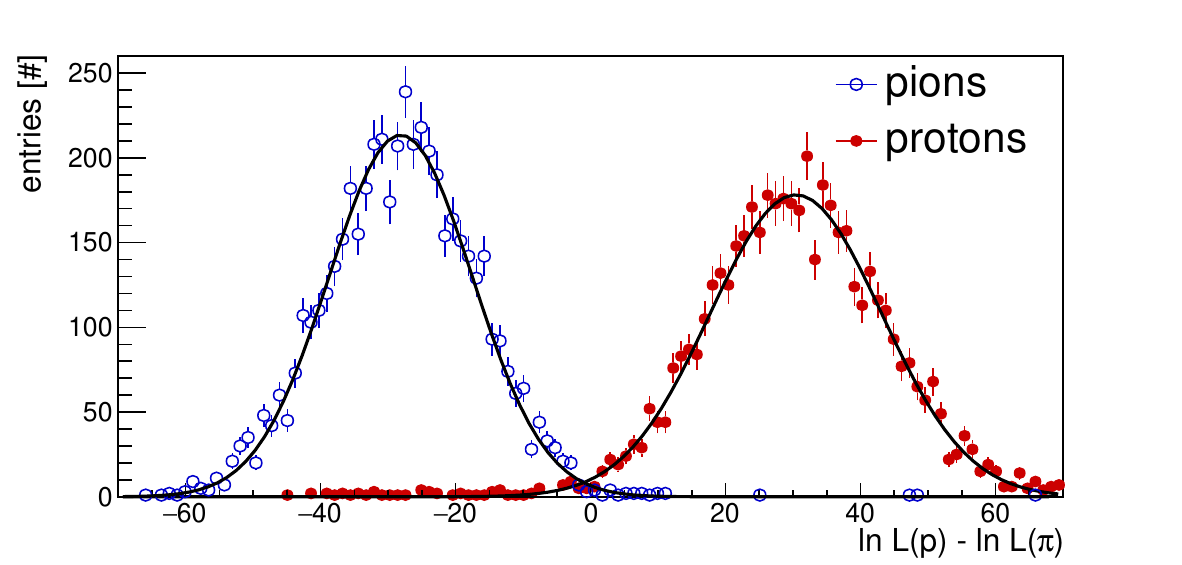}
  \caption{Maximum likelihood difference as a result of time imaging reconstruction for 5000 $\pi$ and $p$ at 7~GeV/$c$ and $20^\circ$ polar angle. Resulting separation power is $5.0 \pm 0.2$ s.d.}
  \label{lh_diff}
\end{figure}

An example of the accumulated hit pattern for 2000 pions at 7~GeV/$c$ momentum and 20$^\circ$ polar angle
is shown in Fig.~\ref{hp} in comparison to Geant4 simulation.
The complexity of the hit pattern is a result of the Cherenkov ring being folded multiple times inside the EV.
The best PID performance was obtained using the time imaging method,
where the measured arrival time of Cherenkov photons in
each single event is compared to the expected photon
arrival time for every pixel and for every particle
hypothesis, yielding the PID likelihoods.
The normalized expected photon arrival time distributions are used as probability
density functions (PDFs) in the likelihood calculations.
A statistically independent sample of events was used to obtain the PDFs.
The maximum likelihood difference obtained for the  most demanding
forward region of $20^\circ$ for $\pi/p$ at 7~GeV/$c$ can be seen in Fig.~\ref{lh_diff}.
The resulting separation power, in this case, is 5.0~$\pm$~0.2 s.d., corresponding to 5.2~$\pm$~0.2 s.d. $\pi/K$
separation at 3.5 GeV/c momentum which satisfies the PANDA Barrel DIRC requirement.

\begin{figure}[ht]
  \centering
  \includegraphics[width=0.9\columnwidth]{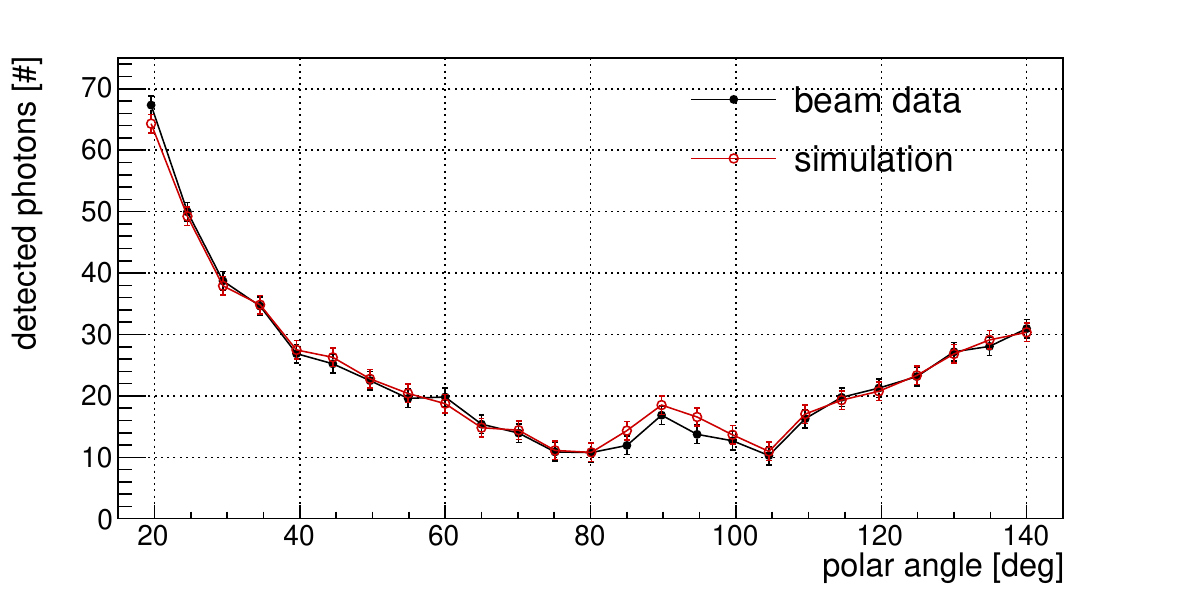}
  \includegraphics[width=0.9\columnwidth]{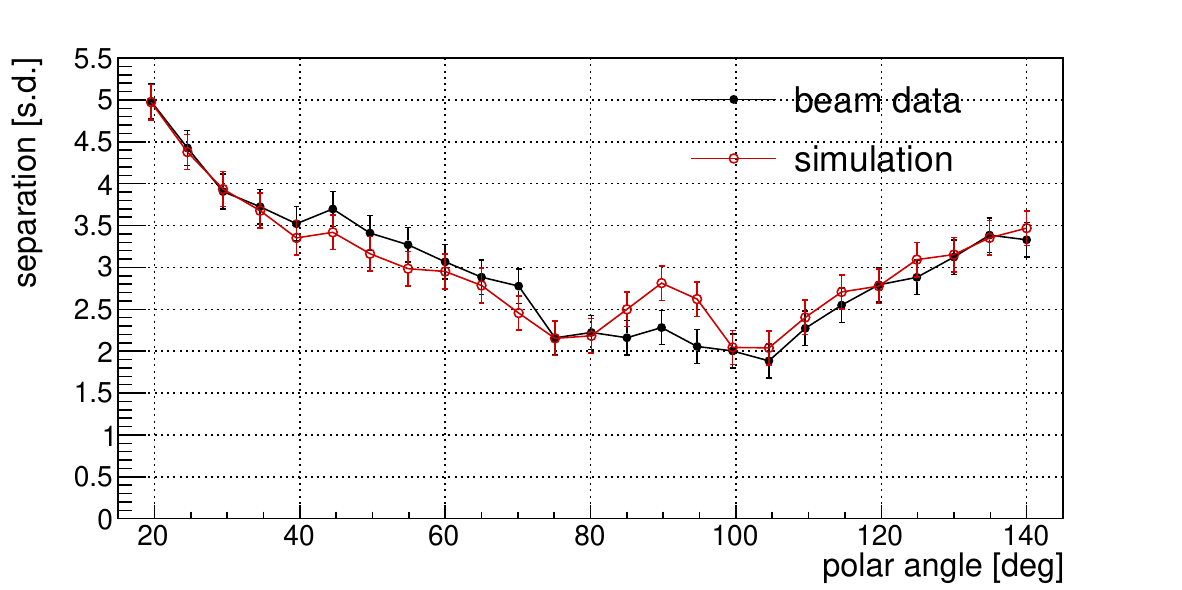}
  \caption{An example of the performance of the PANDA Barrel DIRC prototype at the CERN PS in 2018:
    Number of detected photons (top) and $\pi/p$ separation power (bottom)
    as a function of the beam polar angle at 7~GeV/$c$ momentum for simulation and experimental data.}
  \label{lh_data}
\end{figure}

Fig.~\ref{lh_data} shows the observed performance for the
Barrel DIRC prototype as function of the polar angle at 7~GeV/$c$ momentum.
The number of Cherenkov photons produced by tagged protons ranges from 12 to 68
and is in a good agreement with the simulation.
The photon yield in PANDA is expected to be higher since the new MCP-PMTs
will have a significantly higher photon detection efficiency.

\section{Component production}

\begin{figure}[ht]
  \centering
  \begin{minipage}[t]{.465\linewidth} 
    \includegraphics[width=1\columnwidth]{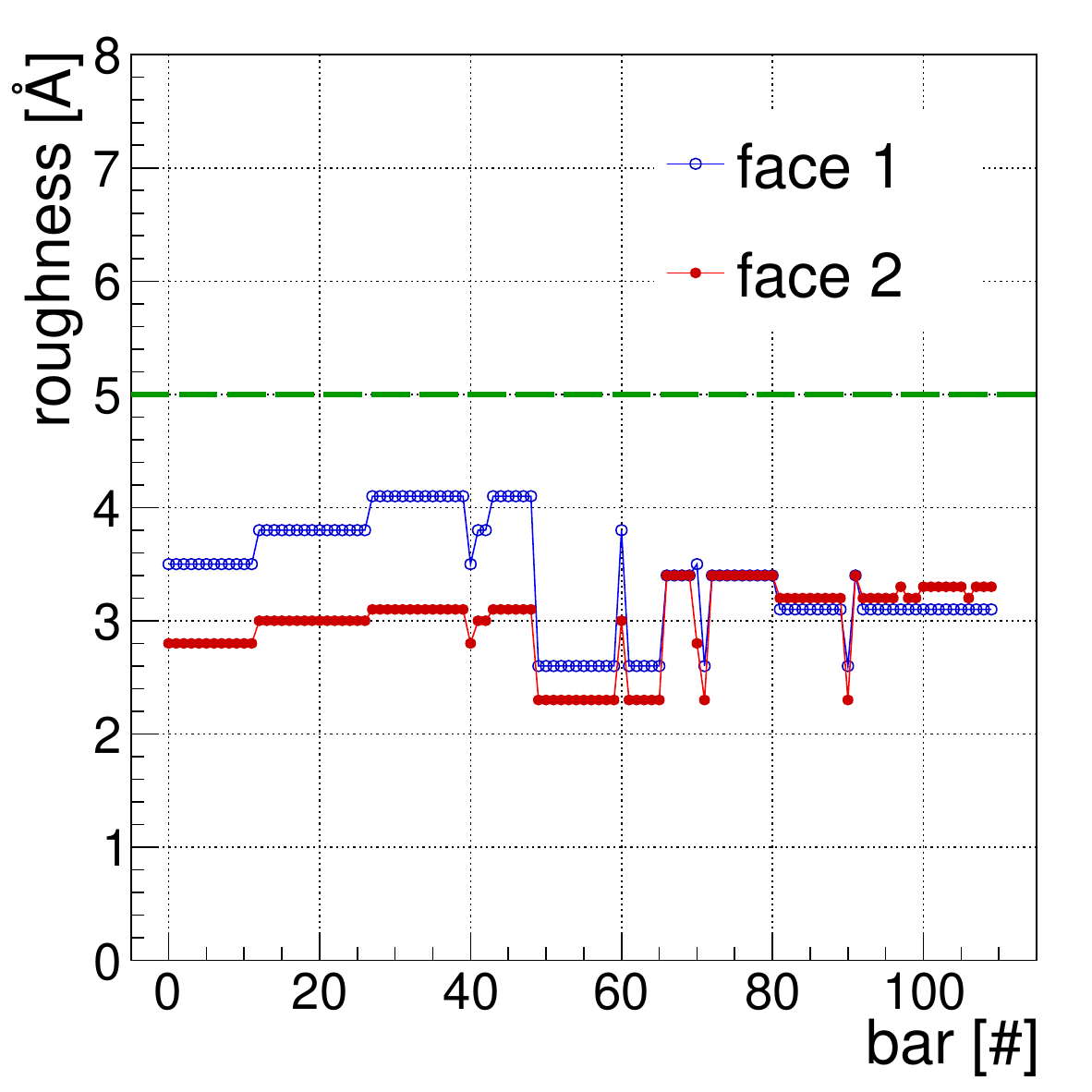}
  \end{minipage}
  \begin{minipage}[t]{.51\linewidth} 
    \includegraphics[width=1\columnwidth]{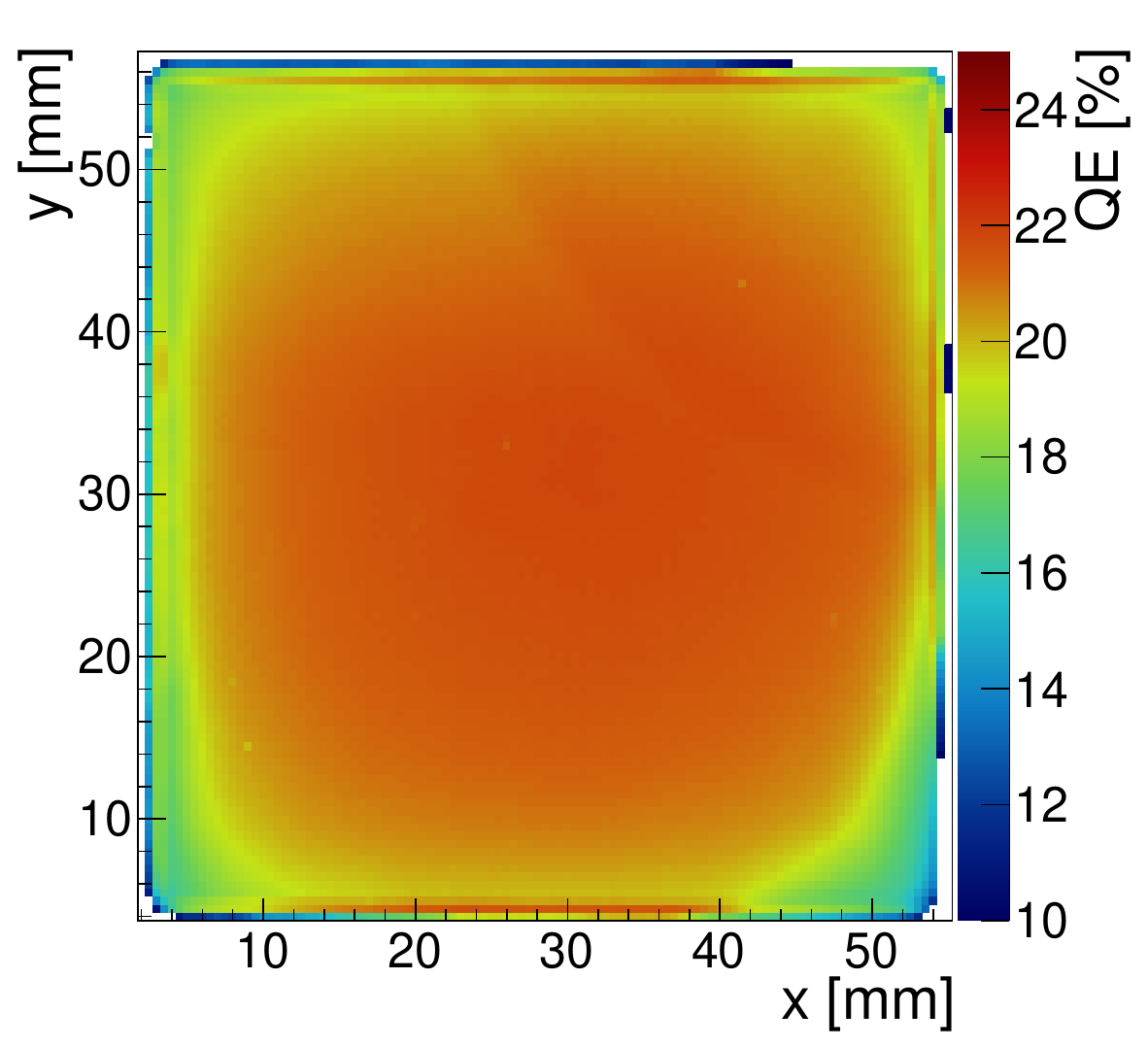}
  \end{minipage}

  \caption{An example of the QA measurements. The left plot shows the surface roughness of the radiator bars
    measured by Nikon Corp. The green line shows the specification requirement for the roughness.  
    The right plot shows the result of a 2D scan of the quantum efficiency of one MCP-PMT measured
    at Erlangen University.}
  \label{qa}
\end{figure}

\begin{figure}[ht]
  \centering
  \includegraphics[width=0.85\columnwidth]{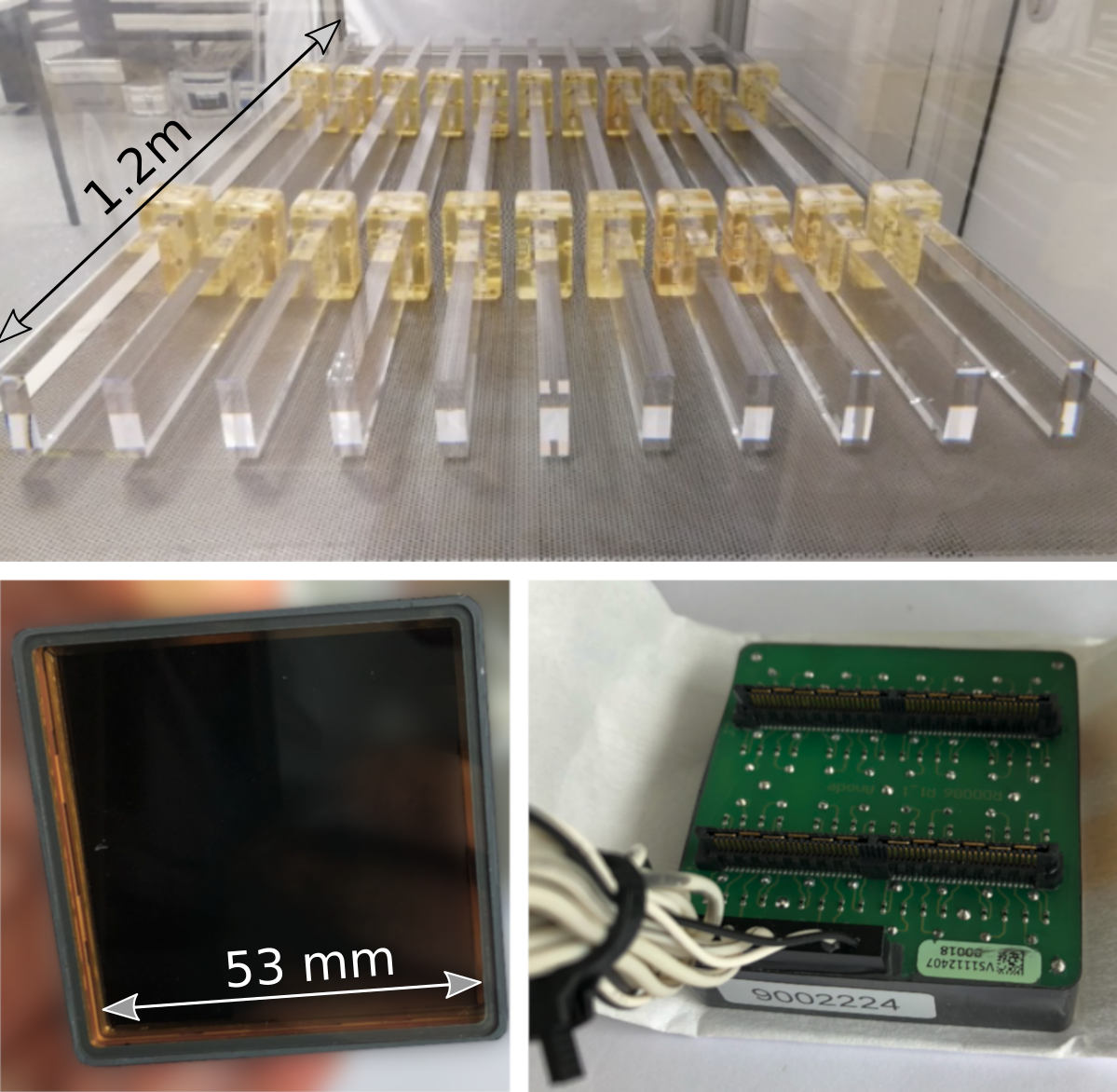}
  \caption{Photographs of the Nikon bars at GSI (top) and
    PHOTONIS Planacon MCP-PMT at Erlangen University (bottom).}
  \label{optics}
\end{figure}

The series production of the PANDA Barrel DIRC components started in 2020.
The contract for the fabrication of the DIRC bars was awarded to Nikon Corp.,
Japan. Between Feb. 2020 and March 2021 a total of 112 bars of 17 $\times$ 53 $\times$ 1200~mm$^3$ size
were delivered to GSI (Fig.~\ref{optics}, top).
96 of them will be glued end-to-end to create the required 48 long bars.
The vendor provided detailed quality assurance (QA) data including  squareness and roughness measurements
(see Fig.~\ref{qa} left).
All parameters are well within the production tolerances.
Additional measurements of the internal reflection coefficient are underway at GSI \cite{pandaqa}.
The series production of the MCP-PMTs (XP85112-S-BA) at Photonis Netherlands BV started in 2021
and the first units were delivered in May 2022 (see Fig.~\ref{optics}, bottom).
The QA measurements for the first MCP-PMTs are performed at Erlangen University
and described in detail in Ref.~\cite{pandamcp}.
Those include measurements of collection efficiency, quantum efficiency and gain homogeneity.
An example of the quantum efficiency measurement is shown in Fig.~\ref{qa} (right).
The next components scheduled for series production are the lenses and prisms.

\section{Conclusions}

The PANDA Barrel DIRC was designed to provide hadronic PID in the target region of the PANDA detector.
Covering the polar angle region of 22-140 degrees it will provide a $\pi/K$ separation power
of at least 3 s.d. for charged particle momenta up to 3.5 GeV/c.
The final design features narrow radiators
made of synthetic fused silica, focusing optics with 3-layer spherical lenses and
a compact prism-shaped expansion volume instrumented with MCP-PMTs.
The latest prototype tests with particle beams at CERN validated this design.
The PANDA Barrel DIRC project has turned from design to component production.
The radiator bars have already been produced and delivered.
The series production of the MCP-PMTs has started. 

\section*{Acknowledgement}

This work was supported by BMBF, HGS-HIRe, HIC for FAIR, and HFHF.
We thank the CERN staff for the opportunity to use the beam facilities
and for their on-site support.

\end{document}